# Fatigue behaviors and atomic-scale mechanisms in nanocrystalline gold thin film

Honglei Chen[1,2], Susheng Tan[4], Zhijie Wang[3]

[1] *Shenyang National Laboratory for Materials Science, Institute of Metal Research, Chinese Academy of Sciences, 72 Wenhua Road, Shenyang 110016, China*

[2] *School of Materials Science and Engineering, University of Science and Technology of China, Shenyang 110016, China*

[3] *Department of Industrial Engineering, Swanson School of Engineering, University of Pittsburgh, PA 15261, United States of America*

[4] *Petersen Institute of Nanoscience and Engineering, University of Pittsburgh, Pittsburgh, Pennsylvania 15260, USA*

**ABSTRACT**

The fatigue properties of 930 nm-thick Au films and 1 μm-thick Au film with a Ti interlayer are systematically investigated. The dominant damage behaviors of 930 nm-thick Au films under dynamic bending cyclic loading changed from extrusions to intergranular cracks with the decrease in strain ranges and the increase in cyclic cycles. The different fatigue behaviors are attributed to the process of edge dislocation annihilation and vacancy formation during cyclic deformation. Depositing 10 nm-thick Ti interlayers between the PI substrates and 1 μm-thick annealed Au films is effective to suppress strain localization and increase the rupture strain and the fatigue properties of thin Au films. This study shed lights on the fatigue mechanism and provide clues to design nanocomposites in the flexible displays in the practical application.

# 1. Introduction

In recent years, flexible electronic devices with high performance, portability, and intelligence have become a hot research area[1, 2, 3, 4, 5, 6, 7]. In practical applications, flexible electronic devices may undergo reciprocating mechanical deformation. For example, wearable electronic devices often experience repetitive bending, twisting, and stretching deformations in daily use [2, 7, 8, 9]. The metal thin film, as a component with metal interconnections within electronic devices, is one of the most vulnerable parts to resist reciprocating mechanical deformation [10]. Due to limitations in the thickness direction of the film, the metal thin film subjected to cyclic deformation on the substrate exhibits fatigue behavior completely different from bulk metal materials[11, 12]. A thorough understanding of the fatigue behavior and failure mechanisms of metal thin films on flexible substrates is not only crucial for building high-performance and long-term reliable flexible circuits but also remains one of the unresolved scientific issues. Over the past two decades, research on metal thin films under cyclic loading has mainly focused on fatigue reliability assessment methods applicable to metal thin films [13, 14, 15], fatigue performance [16, 17], microstructure instability [18, 19, 20, 21, 22, 23, 24], and scale effects and related fatigue damage behavior and mechanisms [25, 26, 27, 28, 29, 30]. A significant amount of research indicates that the formation of voids at the film/substrate interface during the fatigue loading process greatly affects the fatigue performance of metal thin films [28, 31, 32, 33]. Suppressing the formation of voids at the film/substrate interface is a key factor in improving the fatigue performance of metal thin films. Currently, there is limited research on how to design thin film materials to suppress the formation of voids at the film/substrate interface.

In this study, we conducted dynamic bending fatigue experiments on 930 nm thick as-deposited gold films and 1 μm thick annealed gold films with a 10 nm thick titanium bonding layer. We systematically studied the fatigue damage behavior of the gold films. By quantitatively characterizing the extrusion height and transgranular crack density of the gold films after fatigue, combined with the observed surface damage morphology and microstructure characterization, we determined the influence patterns of applied strain range, cycle cyles, and heterointerface on fatigue damage behavior during the fatigue process.

## 2. Methods

### 2.1 Au thin film preparation

Using DC magnetron sputtering, a gold thin film with a thickness of 930 nm was deposited on a 125 μm thick polyimide substrate (Dupont Kapton®). The deposited gold film in the as-deposited state exhibited an out-of-plane {111} texture, with an average grain size of 102 nm ± 28 nm. The deposition of a 1 μm thick gold film on a 125 μm thick polyimide substrate (Dupont Kapton®) was carried out using DC magnetron sputtering under high vacuum conditions (<$10^{-5}$ mbar), with the gold target material purity of 99.99%. Prior to film deposition, argon ion sputtering was employed to clean the polyimide substrate, removing surface contaminants, and enhancing the adhesion between the activated substrate and the film material. To improve the adhesion between the gold film and the polyimide substrate, a 10 nm thick titanium bonding layer was deposited on the substrate. Subsequently, the gold film was deposited on the substrate at a deposition rate of 0.54 nm/s, with a sputtering power of 100 W. Argon gas was used as the protective gas at a flow rate of 30 sccm. The prepared gold film underwent annealing directly in the magnetron sputtering system by heating the sample holder under vacuum conditions. The annealing temperature was 300 ºC, and the annealing time was 30 minutes. The annealed 1 μm thick gold film exhibited an out-of-plane {111} texture, with an average grain size of 120 nm ± 30 nm.

2.2 Mechanical test

The tensile experiments for the two types of gold thin film samples were conducted on a micro-force tensile testing machine. The as-deposited gold thin film, with a thickness of 930 nm, used for the tensile experiments had a gauge section with dimensions of 10 mm × 2 mm. The sample for the tensile experiment of the annealed gold thin film, with a thickness of 1 μm, was in a strip shape with a gauge section measuring 10 mm × 2 mm. The tensile experiments were performed using constant displacement control, with a strain rate of $6.3 \times 10^{-4}$ s$^{-1}$. Metallic thin film materials deposited on flexible substrates are constrained by the substrate, and their mechanical failure modes differ from those of bulk metal materials. Due to the excellent elastic properties of polyimide substrates (fracture strain up to 80%), the metal film usually fractures before the substrate. Therefore, the fracture strain of the film/substrate specimen cannot reflect the true fracture strain of the metal film[15, 34]. Additionally, the failure of the fracture of the metal film is difficult to directly judge with the naked eye.

Generally, the resistance of metal materials is related to the geometric shape, dimensions, and microstructure of the material. Therefore, the resistance method is used during tensile experiments to determine the critical failure point of the material. The resistance value of the metal film specimen increases with the increase in applied strain during uniaxial tensile testing. During the tensile experiment, the relative resistance change varies with applied strain. As strain further increases and reaches the critical failure strain point, the relative resistance changes deviates from the ideal curve. In uniaxial tensile experiments, the strain at which the relative resistance change deviates 5% from the ideal curve is defined as the critical fracture strain[15, 35, 36]. The resistance measurement method used in this experiment is the two-point method, and the measurement equipment is the Agilent 34410 digital multimeter, with a resolution of $10^{-3}$ Ω and a data acquisition interval of 0.1 s. In the mechanical experiment, the resistance meter records the relative resistance change of the metal film in real-time, thus calculating the critical failure point.

In the dynamic bending fatigue experiment, the deposited gold thin film specimens are selected to be strip-shaped, with dimensions of 11 mm × 3 mm. The marked section has dimensions of 6 mm × 3 mm. The annealed gold thin film/substrate specimens, also strip-shaped, have dimensions of 11 mm × 3 mm, with the marked section measuring 6 mm × 3 mm, as illustrated in Figure 3.4b. The dynamic bending fatigue experiment is conducted on a self-built dynamic bending fatigue testing platform[37, 38]. The loading frequency used in the dynamic bending fatigue experiment is 50 Hz, with a loading deflection ranging from 1 mm to 4 mm.

2.3 Material characterization

The extrusion height of the surface of fatigue-affected gold thin film samples was characterized using a laser confocal microscope (OLYMPUS LEXT OLS 4000, Olympus Corporation, Japan). The surface damage morphology of the two types of gold thin film samples after fatigue was characterized using a field emission scanning electron microscope (SEM; Zeiss Supra 35, Carl Zeiss Microscopy, England) at an accelerating voltage of 20 kV. Cross-sectional cutting and observation of the 1 μm thick annealed gold thin film after fatigue were performed using a helium ion microscope (HIM, Zeiss Orion NanoFab, Carl Zeiss Microscopy, England). The microstructure of the fatigue-affected gold thin film was characterized using a transmission electron microscope (TEM; FEI Tecnai F20, FEI, America). The transmission electron microscopy sample preparation method involved initially grinding the sample to approximately 30 μm thickness, followed by ion thinning from the substrate side at -100 ℃ using a precision ion polishing system (PIPS; Model 691, Gatan Incorporated, America).

## 3. Results

3.1 Tensile and fatigue properties of nanocrystalline thin gold film

Figure 1 shows the curves of relative resistance change versus tensile strain for a 930 nm thick deposited gold thin film and a 1 μm thick annealed gold thin film. The black line in the figure represents the predicted ideal curve of relative resistance change with tensile strain. The critical fracture strains for the two thicknesses of gold thin films can be obtained. The critical fracture strain for the deposited gold thin film is 9.4%, while for the annealed gold thin film, it is 21%. Figure 2 shows the fatigue life curves as a function of applied cyclic strain range for a 930 nm thick deposited gold thin film and a 1 μm thick annealed gold thin film. From the figure, it can be observed that the fatigue performance of the annealed gold thin film is superior to that of the deposited gold thin film. Particularly in the high-cycle region, the fatigue performance of the annealed gold thin film is significantly better than that of the deposited gold thin film. A significant amount of research indicates that the strength and fatigue performance of metal thin films increase with a reduction in film thickness/grain size [18, 26, 28]. In comparison to the 1 μm thick annealed gold thin film, the 930 nm thick deposited gold thin film has smaller film thickness and grain size (the average grain size of the deposited gold thin film is 102 nm, while the grain size of the annealed gold thin film is 120 nm). Theoretically, the fatigue performance of the deposited gold thin film should be superior to that of the annealed gold thin film; however, in practice, the fatigue performance of the two gold thin films is opposite to the theoretical results.

3.2 Surface morphology of the deformed thin Au films

Figure 3 shows scanning electron microscope (SEM) observations of the annealed gold thin film after tensile testing. From the figure, it can be observed that the primary failure mode of the gold thin film is the nucleation and extension of microcracks. The surface buckling is attributed to the modulus mismatch between the gold thin film and the polyimide substrate. The substrate contraction induces compressive stress perpendicular to the direction of the cracks in the gold thin film, causing the already cracked gold thin film to delaminate and protrude. The damage morphology of the annealed gold thin film after tensile testing is similar to that reported for copper thin films [12] and silver thin films [39] in the literature.

Figure 4 illustrates the surface damage morphology of the 930 nm thick deposited gold thin film and the 1 μm thick annealed gold thin film. In both types of films, fatigue damage resembling the intrusions/extrusions observed in bulk metallic materials can be observed, as shown in Figures 4a and 4b. This fatigue damage behavior is very similar to the fatigue damage behavior reported in copper thin films [11, 14, 31, 40, 41, 42] and silver thin films [31] in the literature. Fatigue cracks primarily initiate at the extrusions and propagate during subsequent cyclic loading processes. Additionally, it can be observed from the figures that the fatigue cracks are not typical Type I (opening) cracks. There is a slight deviation in the propagation of fatigue cracks, and the crack propagation path is roughly perpendicular to the loading direction. Differently, in the high-cycle region, the fatigue damage behavior of the deposited gold thin film includes not only fatigue extrusions but also a significant number of transgranular cracks, as shown in Figure 4c. However, under the same failure cycles, the surface of the annealed gold thin film still exhibits fatigue extrusions without observable transgranular cracks, as depicted in Figure 4d. Furthermore, in the high-cycle region at a low cyclic strain range (0.14%), the surface of the deposited gold thin film after fatigue only shows transgranular cracks, as seen in Figure 4e. However, under the same conditions, the surface of the annealed gold thin film does not exhibit obvious fatigue damage, as shown in Figure 4.

To recap, at high applied cyclic strain ranges, the only fatigue damage behavior observed on the surface of the gold thin film is extrusion damage. As the applied cyclic strain range decreases and the cycle count increases, the fatigue damage behavior gradually transitions from extrusion damage to both extrusion and transgranular crack damage. At low applied cyclic strain ranges and high cycle counts, the surface of the gold thin film after fatigue only exhibits transgranular crack damage. Therefore, it can be inferred that the extrusion damage behavior and transgranular crack damage behavior in the fatigue process of the deposited gold thin film compete. As the applied cyclic strain range decreases and the cycle count increases, the fatigue damage behavior shifts from extrusion damage to transgranular crack damage.

Figure 5 shows cross-sectional observations of extrusion sites in the deposited gold thin film after fatigue. It is clear from the figure that pores form at the film/substrate interface of the deposited gold thin film after fatigue. The cross-sectional morphology of the deposited gold thin film after fatigue is similar to that reported for copper thin films [11, 28, 32] and silver thin films [31] in the literature. Numerous studies have indicated that metal thin film/substrate specimens develop pores at the film/substrate interface after fatigue, and subsequent fatigue cracks initiate and propagate at these pores [11, 28, 31, 33]. In contrast to the deposited gold thin film, to enhance the adhesion between the film and the substrate, we introduced a 10 nm thick titanium bonding layer between the annealed gold thin film and the substrate. To investigate the influence of the titanium bonding layer on fatigue damage behavior in the annealed gold thin film, helium ion microscopy was used to cut and observe the cross-sectional morphology at extrusion damage sites after fatigue, as shown in Figure 6. It can be seen that there are no apparent pores present at the film/substrate interface. This experimental observation differs from the cross-sectional morphology of the deposited gold thin film.

Observations of the deposited gold thin film after fatigue using transmission electron microscopy (TEM) (Figure 7) reveal a significant presence of dislocations within the grain interiors, as shown in Figure 7a. The dislocation configurations are predominantly single dislocations or entangled dislocations, with no evidence of cell-like or wall-like dislocation structures forming inside the grains. No specific dislocation structures were observed near transgranular cracks, as shown in Figure 7b. Further high-resolution transmission observations of the deposited gold thin film after fatigue are presented in Figure 8. Figure 8 depicts high-resolution TEM images of the deposited gold thin film after fatigue. From the figure, it can be seen that numerous dislocation dipoles appear within the grain interiors after fatigue (highlighted by yellow circles in the image)[43, 44, 45, 46, 47, 48, 49].

## 4. Discussions

4.1 Fatigue mechanism

The fatigue damage failure of metal thin films can be specifically divided into the following three processes: firstly, surface roughening and the formation of voids at the film/substrate interface; subsequently, fatigue cracks initiate at these voids and propagate towards the surface; finally, the fatigue cracks continue to expand to form observable patterns [28, 33, 50]. Many scholars have observed experimental phenomena that confirm this series of damage evolution processes. Kraft *et al*. [32] observed that in some grains forming extrusions, only voids appeared without fatigue cracks. Fatigue cracks could only be observed when both voids and extrusions appeared in the film. Thus, voids are crucial in the fatigue process of metal thin films. The time required to form a sufficiently large void at the film/substrate interface (the time for void initiation and growth) determines the fatigue life of the metal thin film. The number of vacancies generated during the fatigue process of metal thin films is jointly determined by the film thickness, grain size, and applied cyclic plastic strain range. During the fatigue experiment, dislocation multiplication and motion continually occur within the grains of the metal thin film. Prolonged dislocation motion through the film leaves a segment of mismatched dislocations at the film/substrate interface. Additionally, during cyclic loading, dislocation loops with opposite sign Burgers vectors continually form within the grains. The annihilation of these loops with opposite sign dislocations leads to the formation of oversaturated vacancies at the film/substrate interface.

For a 1 μm thick metal film, at a plastic strain range of 1%, the vacancy concentration is 0.005. This value far exceeds the vacancy concentration at the melting point of face-centered cubic metals (0.001) [51]. Therefore, according to the proposed mechanism, the metal film can generate a sufficient number of vacancies during the fatigue loading process for the initiation and subsequent growth of voids. Kraft *et al*. [32] conducted cross-sectional observations of fatigued copper film/substrate specimens and found voids forming near the film/substrate interface. Ruth *et al*. [11,31] also found the presence of voids at the film/substrate interface in fatigued copper and silver films. As the number of cycles gradually increases, the vacancy concentration continues to increase. Since vacancies formed at the film/substrate interface diffuse to the grain boundaries, the diffusion process causes the vacancy concentration at the intermediate position to be higher than at the two sides. After a certain number of cycles, when the local vacancy concentration reaches the critical vacancy concentration, oversaturated vacancy aggregation forms voids. Therefore, during the fatigue loading process, the number of cycles and the applied cyclic strain range influence the process of vacancy aggregation, leading to the formation of voids, thereby affecting the fatigue damage behavior of the metal film, and determining its fatigue life.

4.2 The effect of applied strain range and cyclic cycles on the fatigue damage behaviors.

The fatigue damage behavior controlled by the applied cyclic strain range and the number of cycles can be explained using the vacancy formation mechanism described in Section 4.1. The number of vacancies generated in half a cycle depends on the applied cyclic strain range. When the applied cyclic strain range is large, a large number of vacancies are generated in half a cycle, and the vacancy generation rate is high. This leads to a rapid accumulation of local vacancies with a high concentration, making it easy for the local vacancy concentration to reach the critical vacancy concentration, forming voids. When the applied cyclic strain range is low, and the number of cycles is large, the vacancy formation rate is low, and the accumulation of local vacancies is slow with a low concentration. Many vacancies diffuse to the surface and grain boundaries. The large number of vacancies diffusing to the grain boundaries leads to the formation of transgranular cracks. The vacancies originally intended for void formation are ultimately used to form transgranular cracks. There is a competitive relationship between the vacancies intended for void formation and those intended for transgranular crack formation, so both void extrusion and transgranular cracks can be observed in this region. The specific fatigue damage behavior occurring in the metal film depends on the diffusion distance (grain size), diffusion time (number of cycles), and vacancy generation rate (applied cyclic strain range). Therefore, we can observe the extrusion height and transgranular crack density determined by the number of cycles and applied cyclic strain range. The more vacancies used for extrusion (larger applied cyclic strain range and more cycles), the higher the extrusion height on the surface of the fatigued film.

Similarly, the higher the number of vacancies used for transgranular crack formation (smaller applied cyclic strain range and more cycles), the higher the density of transgranular cracks after fatigue. Due to the competitive relationship between the two, we can see that even with a large accumulated cyclic strain, the extrusion height is small. This is because the applied strain range is small, causing the vacancies originally intended for extrusion to diffuse to the grain boundaries and surface, reducing the vacancies used for extrusion and decreasing the extrusion height.

4.3 The influence of Ti interlayer on the mechanical behaviors of thin Au film

Generally, the mechanical properties of metal films follow the trend of becoming stronger as they become thinner. As the thickness of the metal film increases, the yield stress decreases [12, 15, 18]. The reduced yield stress will decrease the overall stress level in the metal film during the tensile process, making it more difficult for the film to delaminate from the substrate [12]. Due to the larger thickness and grain size of the annealed gold film, the critical fracture strain of the annealed gold film is higher than that of the deposited gold film. Additionally, in the annealed gold film/substrate sample, there is a 10 nm thick titanium bonding layer between the gold film and the substrate. The presence of the 10 nm thick titanium bonding layer enhances the interface bonding strength, allowing the gold film to adhere better to the substrate. The increased interface strength can more effectively suppress the process of strain localization during cyclic loading, thereby delaying and suppressing the occurrence of fracture failure behavior [52]. Compared to the deposited gold film, the annealed gold film has a more stable microstructure after annealing treatment and is less prone to grain growth behavior under mechanical loading. The annealed gold film can maintain the excellent performance derived from the original fine grain size. Therefore, for these reasons, the tensile performance of the annealed gold film is superior to that of the deposited gold film.

During the fatigue process, we can explain why the fatigue performance of the annealed gold film with a 10 nm thick titanium film layer as a bonding layer is superior to that of the deposited gold film. During the fatigue process, vacancies generated by dislocation annihilation can diffuse to the surface, grain boundaries, and the film/substrate interface. For the deposited gold film/substrate sample, the film/substrate interface is not a very effective vacancy annihilation site. During fatigue loading, a large number of vacancies locally accumulate to form voids, and then fatigue cracks initiate and propagate at the voids, ultimately leading to fatigue failure of the metal film. A small number of vacancies generated during the fatigue process will diffuse to the grain boundaries and surface. In contrast to the deposited gold film/substrate sample, the annealed gold film/substrate sample has a 10 nm thick titanium film layer as a bonding layer, introducing an additional titanium film/gold film interface within the sample. This heterojunction interface can effectively act as a vacancy annihilation site during the fatigue process, inhibiting the formation of voids at the film/substrate interface and ultimately slowing down the initiation and propagation of fatigue cracks. Due to the presence of the titanium film layer, no voids were observed near the substrate in the annealed gold film. Therefore, the gold film/titanium film interface enhances the fatigue performance of the annealed gold film. Additionally, as the number of cycles increases, the number of vacancies generated during the fatigue loading process increases, and the annihilation effect of the titanium film/gold film interface on vacancies becomes more pronounced. Therefore, in the high-cycle fatigue region, the fatigue performance of the annealed gold film is significantly superior to that of the deposited gold film. On the other hand, the titanium film bonding layer between the substrate and the gold film can increase the adhesion between the film and the substrate. The increased interface strength will help suppress strain localization, thereby delaying fatigue failure. Considering these two aspects of the elucidated reasons, we can conclude that the presence of the gold/titanium interface makes the performance of the annealed gold film superior to that of the deposited gold film. The reason why no transgranular cracks appeared in the annealed gold film at low applied strain ranges and high cycle numbers can also be attributed to this 10 nm thick titanium film layer. It is precisely because the titanium/gold interface can act as a vacancy annihilation site during the fatigue process, annihiliting vacancies originally intended for void formation and those diffusing to grain boundaries. Therefore, the grain boundaries of the annealed gold film did not annihilate enough vacancies during the fatigue process to form transgranular fatigue damage.

## 5. Conclusions

This study investigates the tensile properties, fatigue performance, fatigue damage behavior, and related mechanisms of 930 nm thick deposited gold thin films on polyimide substrates and 1 μm thick annealed gold thin films with a 10 nm titanium film as a bonding layer through uniaxial tensile tests, dynamic bending fatigue tests, SEM, and TEM characterization techniques. The main conclusions obtained are as follows:

1. Due to the presence of a 10 nm thick titanium bonding layer, the tensile and fatigue performance of the 1 μm thick annealed gold thin film is significantly better than the fatigue performance of the 930 nm thick deposited gold thin film. This improvement is attributed to the increased interface strength suppressing strain localization during mechanical loading. Additionally, the titanium film/gold film interface can act as sites for vacancy annihilation during cyclic loading, inhibiting the formation of voids and subsequent crack initiation.

2. The density of transgranular cracks on the surface of the 930 nm thick deposited gold thin film after fatigue is determined by the applied cyclic strain range and cycle count. With a lower applied cyclic strain range and higher cycle count, the density of transgranular cracks on the surface of the gold film after fatigue increases.

3. As the applied cyclic strain range decreases and the cycle count increases, the fatigue damage behavior of the 930 nm thick deposited gold thin film transitions from extrusion to transgranular cracks, and these two damage behaviors compete. The applied cyclic strain range and cycle count determine the diffusion to grain boundaries and the local aggregation of vacancies, influencing the number of voids formed and thus affecting fatigue damage behavior during the fatigue process.

**Author Contributions**

H.C. conducted the experiments and performed the data analysis. H.C., S.T. and Z.W. prepared the paper with the contribution of all authors.

**Conflict of interest**

The authors declare no competing financial interest.

**Figure Captions**

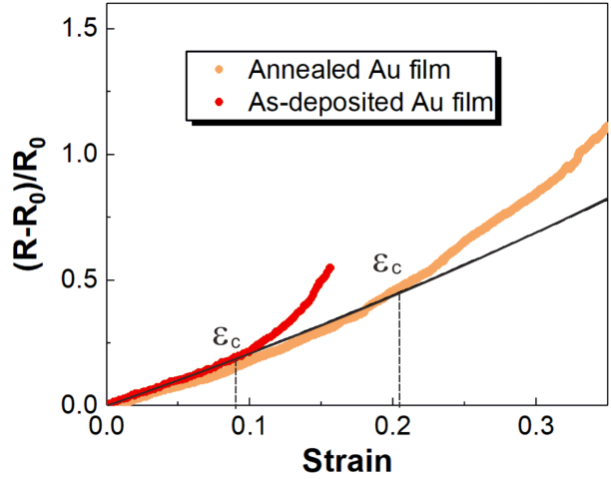

**Figure 1. Relative resistance increase vs applied strain in the Au films during tensile tests.**

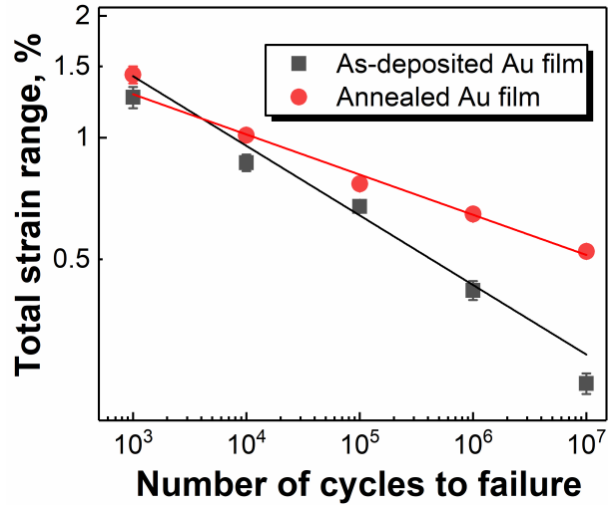

**Figure 2. Applied strain range-fatigue life curve of the as-deposited Au films with a thickness of 930 nm and the annealed Au films with a thickness of 1 μm.**

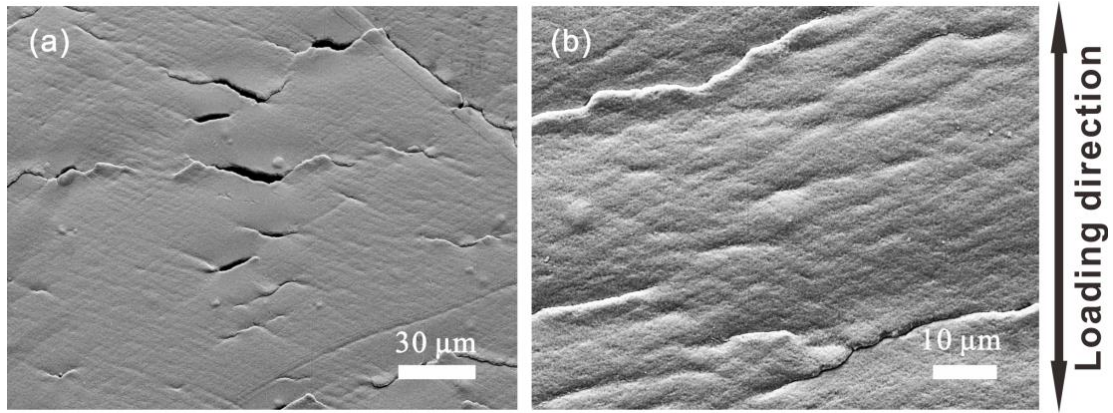

**Figure 3. SEM observation of crack morphology in the 1 μm-thick annealed Au films.**

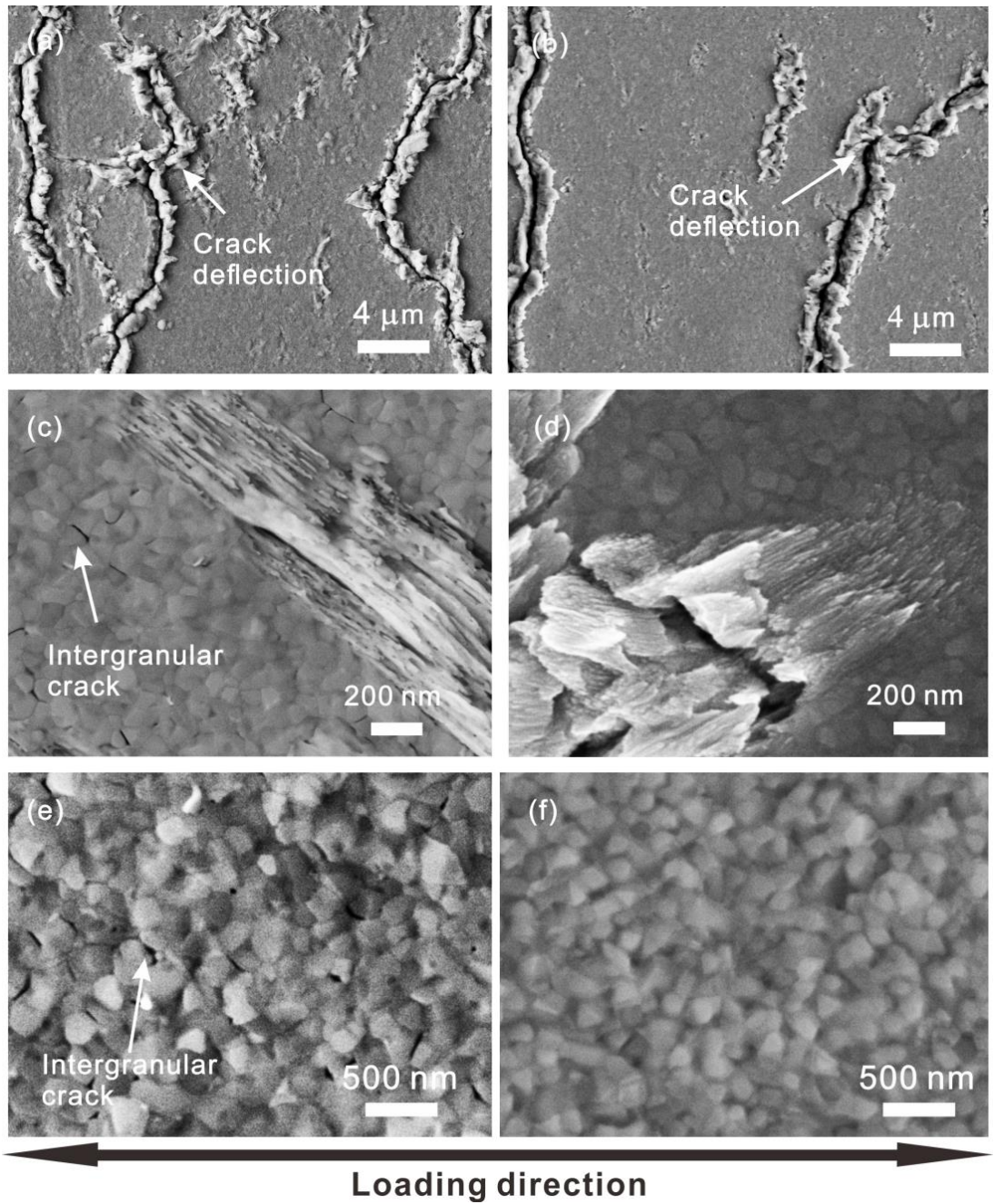

**Figure 4.** SEM observation of damage morphology in Au films with a thickness of (a) 930 nm at $\Delta\varepsilon=1.41$ % and $N_f=10^5$, (b) 1 μm at $\Delta\varepsilon = 1.32$ % and $N_f=10^4$, (c) 930 nm at $\Delta\varepsilon = 0.46$ % and $N_f=10^8$, (d) 1 μm at $\Delta\varepsilon = 0.61$ % and $N_f=10^8$, (e) 930 nm at $\Delta\varepsilon = 0.14$ % and $N_f=10^8$, (d) 1 μm at $\Delta\varepsilon = 0.2$ % and $N_f=10^8$.

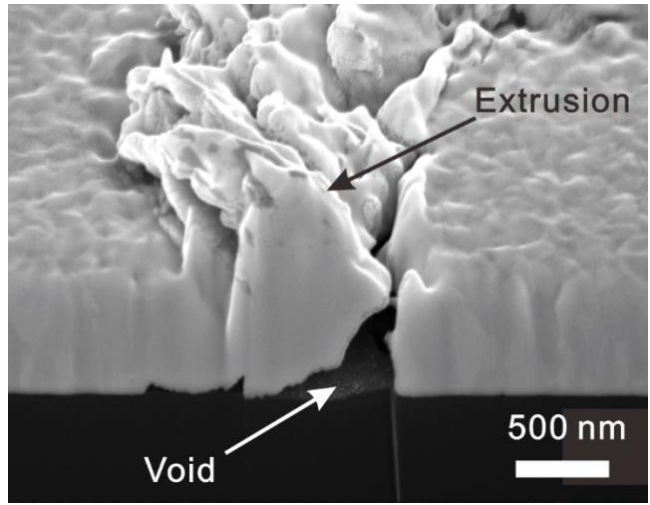

**Figure 5. SEM image of the cross-sectioned extrusions in 930-thick as-deposited Au films after fatigue tests.**

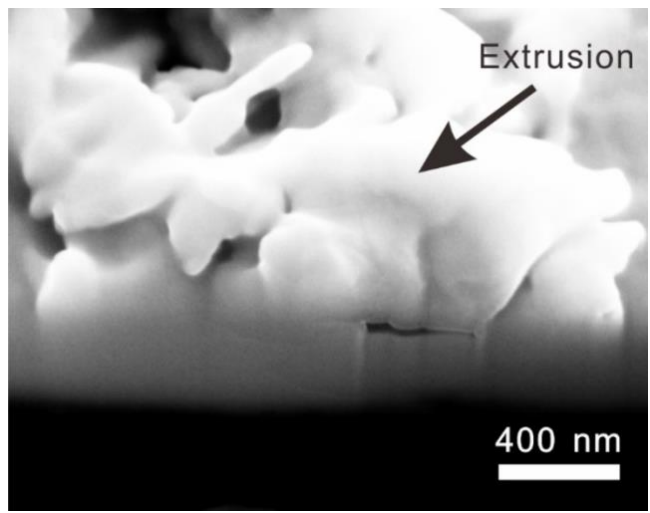

**Figure 6. SEM image of the cross-sectioned extrusions in a fatigued 1 μm-thick annealed Au film after $10^4$ cycles.**

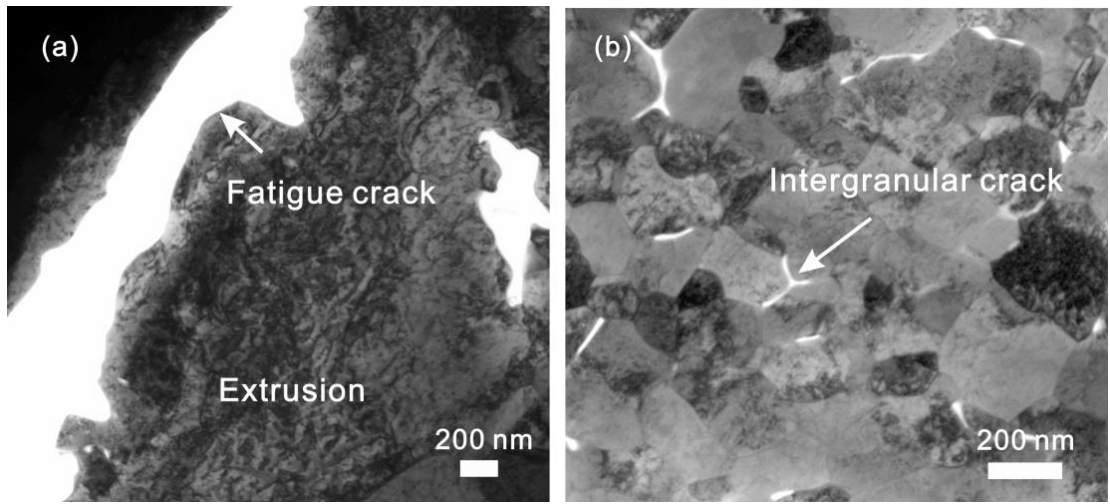

**Figure 7.** TEM images in the fatigued 930 nm-thick Au films (a) around the fatigue cracks, (b) around the intergranular cracks.

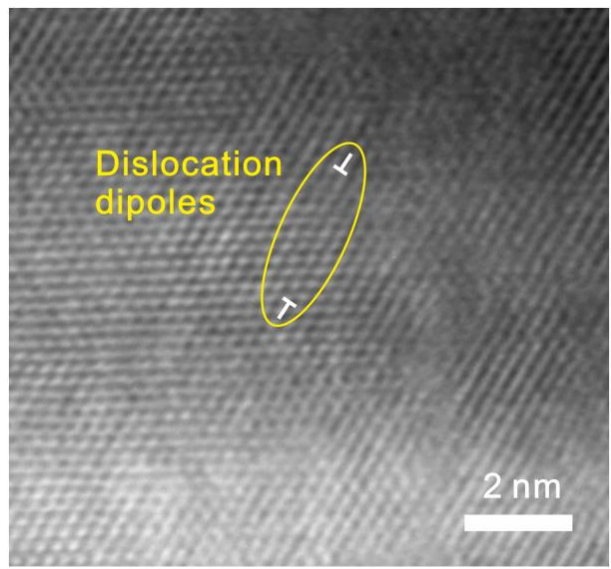

**Figure 8.** HRTEM images of the fatigued 930 nm-thick Au films.